\documentclass[aps,pra,showpacs,floatfix,twocolumn,superscriptaddress]{revtex4}
\usepackage{graphicx}
\usepackage{bm,amsmath}
\usepackage{dcolumn}
\usepackage{mathrsfs}
\usepackage{cases}
\usepackage{color}
 \usepackage{ulem}
 \usepackage{indentfirst}
\newcommand{\be}{\begin{equation}}
\newcommand{\ee}{\end{equation}}
\newcommand{\bea}{\begin{eqnarray}}
\newcommand{\eea}{\end{eqnarray}}
\newcommand{\la}{\langle}
\newcommand{\ra}{\rangle}
\begin{document}
\title{Trimer superfluid induced by photoassocation on the state-dependent optical lattice}
\author{Wanzhou Zhang}
\email{zhangwanzhou@tyut.edu.cn}
\affiliation{College of Physics and Optoelectronics, Taiyuan University of Technology, Shanxi 030024, China}
\affiliation{Key Laboratory of Advanced Transducers and Intelligent Control System, Ministry of Education,
 Taiyuan University of Technology, Shanxi 030024, China}

 \author{Ran  Li}
\affiliation{College of Physics and Optoelectronics, Taiyuan University of Technology, Shanxi 030024, China}
 \author{W. X. Zhang}
\affiliation{College of Physics and Optoelectronics, Taiyuan University of Technology, Shanxi 030024, China}

\author{C. B. Duan}
\affiliation{College of Science, Chang an University, Xi‘an 710064, China}
\author{T. C.  Scott}

\affiliation{College of Physics and Optoelectronics, Taiyuan University of Technology, Shanxi 030024, China}

\date{\today}
\begin{abstract}
We use the mean-field method, the Quantum Monte-carlo method and the Density matrix renormalization
group  method to  study the trimer superfluid phase and the quantum phase diagram of
 the Bose-Hubbard model in an optical lattice, with explicit trimer tunneling term.
 Theoretically, we derive  the explicit trimer hopping terms, such as $a_i^{3\dagger}a_j^3$,
by the Schrieffer-Wolf transformation. In practice, the trimer super\-fluid
described by these terms is driven by photoassociation.
The phase transition between the trimer super\-fluid phase and other phases are also studied.
Without the on-site interaction,  the phase transition between the trimer superfluid phase and the Mott Insulator phase
is continuous. Turning on the on-site interaction, the phase transitions are first order
with Mott insulators of atom  filling $1$ and $2$.
With nonzero atom tunneling, the phase transition is first order from the atom
superfluid  to the trimer superfluid. In the trimer superfluid phase, the  win\-ding numbers
can be divided by three without any remainders. In the atom superfluid and pair superfluid,
the vorticities are $1$ and $1/2$, respectively. However, the vorticity  is $1/3$ for the trimer superfluid.
The power law decay exponents is $1/2$ for the non diagonal correlation $a_i^{\dagger 3} a_j^{3}$,
i.e. the same as the exponent of the correlation  $a_i^{\dagger}a_j$ in hardcore bosons.
The density dependent atom-tunneling term $n_i^2a_i^{\dagger}a_j$
and pair tunneling term  $n_ia_i^{\dagger2}a_j^2$  are also studied.
With these terms,
the phase transition from the empty phase to atom  superfluid  is  first order and
different from the cases without the density dependent terms.
The phase transition is still first order from the trimer superfluid
phase to the atom superfluid.
The ef\-fects of temperature  are  studied.
Our results will be helpful in realizing the trimer superfluid by a cold atom experiment.

\end{abstract}
\maketitle
\section{Introduction.}
\label{sec:intro}
In recent years, the Bose-Hubbard model using an on-site attractive
 interaction\cite{th1,3bodyloss,3body2,mfy1,mfy2,psf,ycc,fintemp},
 or with explicit pair tunneling term\cite{xfzhou,wz1,wz2}  has caught
the attention of theoretical and experimental physicists.
As for a\-toms with a two-body attractive interaction on an optical lattice,
bosons can form an interesting spontaneous pair-tunneling
between two of the nearest neighbor sites.
This motion forms a pair super\-fluid  state (PSF)
which is stabilized by a
three-body constraint, even though the Hamiltonian has no obvious
pair-tunneling term.
The three-body constraint($a^{\dagger 3}=0$),
e.g. two a\-toms allowed in a site,
has been realized by large three-body loss processes\cite{3bodyloss,3body2}.

Besides the attraction-induced pair superfluid, virtually excited atom-pairs driven
by atom-molecule coupling, could lead to a PSF. Compared with a spontaneous PSF,
 this artificially driven super\-fluid covers a lar\-ger parameter area in the phase diagram.
So the PSF is more detectable in experiments and  it is easier to explore the
phase transition between a PSF and other phases\cite{xfzhou,wz1,wz2}.

Naturally, the question arises:
can the trimer super\-fluid(TSF) exist in a stable fashion and be detectable in the real experiments?
In the TSF phase, three atoms on one site will tunnel into the neighborhood sites simultaneously.
An on-site trimer can exist by using a
Bose gas on the optical lattice\cite{1st},
in which
two atoms are in the lowest hyperfine state, and the third atom sits in an excited hyperfine state.
Other Bose trimers have been observed in  the several few-body systems, such as   7Li\cite{li7},  39K\cite{39K}, 85Rb\cite{85Rb} and 133Cs\cite{133Cs}.

Evidence about an Efimov superfluid\cite{efimov,efimovsf} and trimer condensation in a frustrated system\cite{zhaihui} on the Kagome lattice
shows that exploring   the macroscopic behaviors of
trimer tunneling, instead of small clusters,  is  a  interesting topic in the cold atom regime.
Actually, by loading the atoms into a state dependent lattice\cite{xfzhou},
the laser could  drive the pair atoms tunnel on the optical lattice.
In high densities, there is a possibility
to form tri\-mers as the collision occur between dimers and
a\-toms by photoassociation\cite{PA}.
So it is possible to drive the tri\-mers tunnel on the lattice and explore the TSF phase and other kinds of macroscopic behavior.

\begin{figure}[t]
\includegraphics[width=4cm]{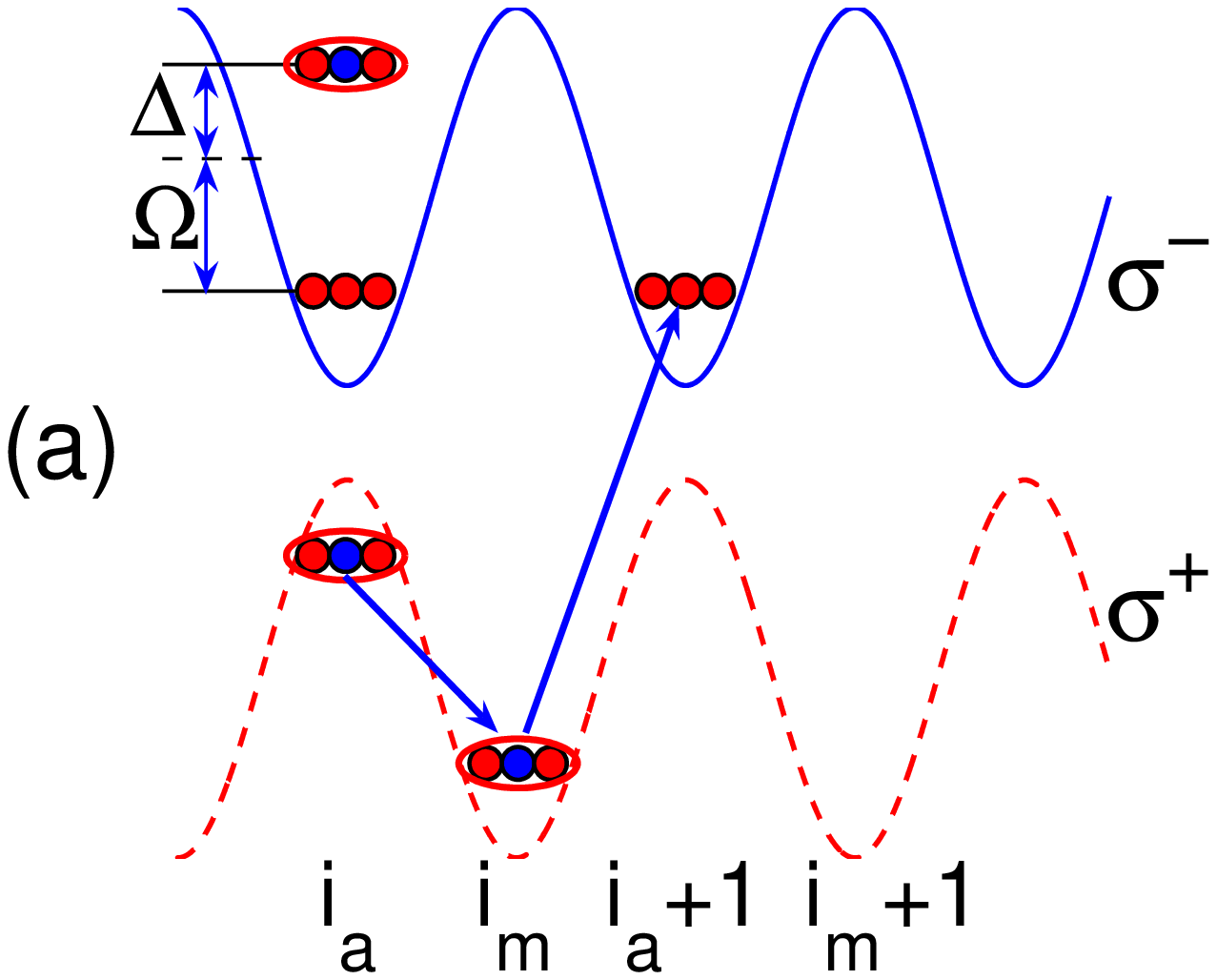}
\includegraphics[width=4cm]{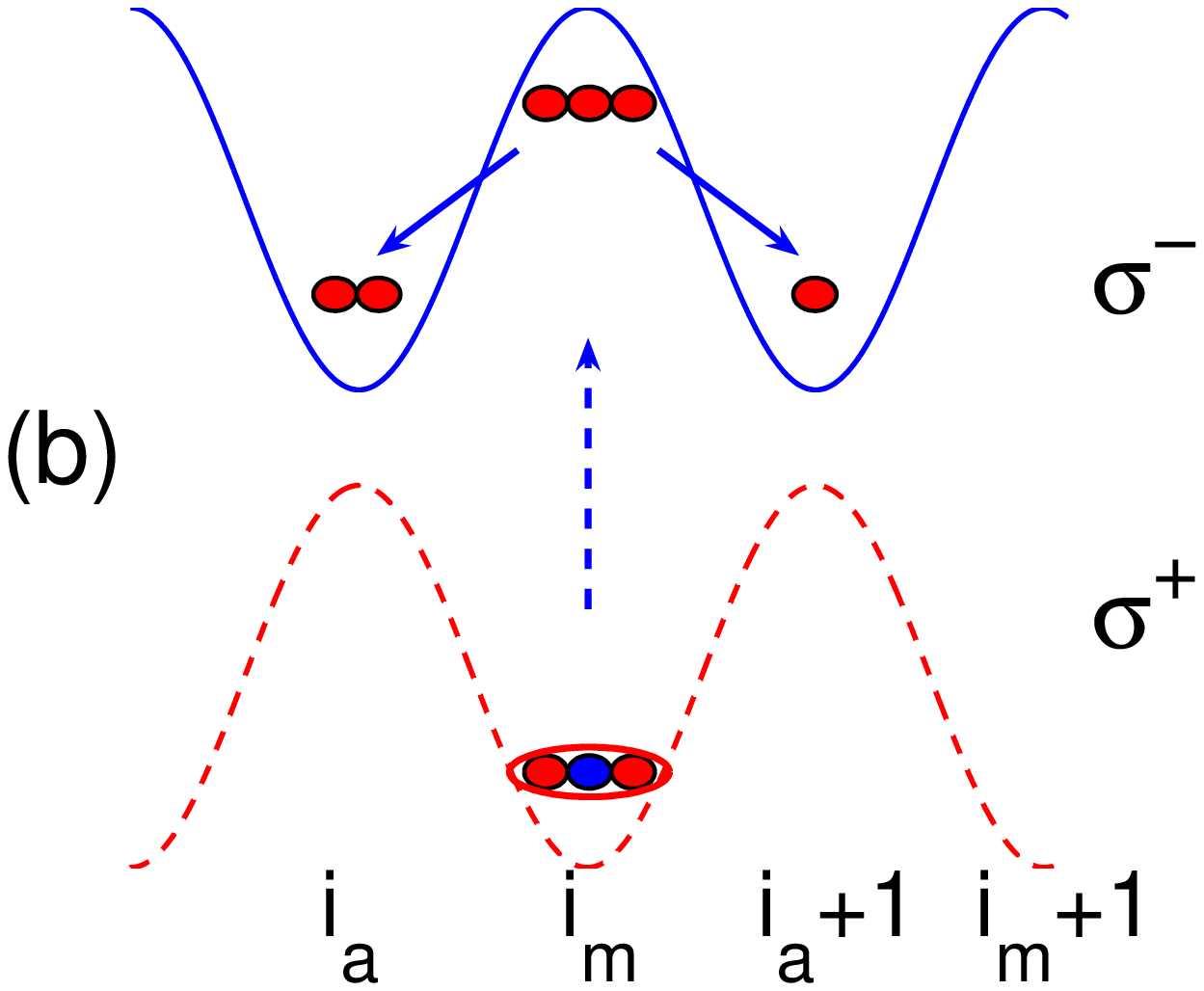}
\caption{(Color online) (a) Experimental realization of trimer hopping
in a state dependent optical lattice.
$\Delta$ is the detuning parameter and  $\Omega$  represents the atom-molecule coupling.
The red and  blue circles represent bosons in the
lowest and (excited) hyperfine states, respectively.
(b) The atom tunneling and pair
tunneling in the procedure of trimer formation.
The probability of a separation is smal\-ler than the probability of the trimer formation.}

\label{ex}
\end{figure}

In the present work,
we provide a model with trimer hopping terms and additional  new terms
analytically and the possibilities of
various ratios of the parameters.
We actively find the TSF on the state-dependent optical lattice, in certain parameter regions.
The phase transition between the trimer super\-fluid and other phases are also
studied.
The typical distribution  of win\-ding numbers are shown with the vorticity of the TSF phase being $1/3$.
The power law decay exponents  of  the correlation are discussed
as well the ef\-fects of density dependent tunneling are studied.
The finite-temperature phase diagrams  are  provided both with and without the atom tunneling.

The outline of this work  is  as follows. Sec.~\ref{sec:model} shows
the derivation of the model, the controllable range of the parameters, and the experimental realization.
We describe the methods and useful observables in Sec.~\ref{sec:method}. We
then present the ground state phase diagram and the TSF phase, in Sec.~\ref{sec:gpt}, for various cases.
The win\-ding numbers and
correlations are also dis\-cussed.
Sec.~\ref{sec:dependent} shows  the ef\-fects of
den\-sity dependent terms. In particular, we show the first-order  transition
between the empty phase and the atom superfluid (ASF) phase.
The first-order transition of ASF-TSF is also shown.
Sec. \ref{sec:finitet} focusses on the finite-temperature properties of the TSF phase.  Concluding
comments are made  in Sec.~\ref{sec:discon}.

\section{The model}
\label{sec:model}

\subsection{Experimental realization of trimer tunneling}

Figure \ref{ex}(a) shows a possible path for  the trimer tunneling on the state dependent lattice.
This scheme is different from the one described in pre\-vious work\cite{1st},
which uses the usual optical lattice.
We make  the initial three free  $^{7}$Li $^2S_{1/2}$  atoms sit in the well $i_a$.
We  excite one of the three atoms  to  state $^2P_{3/2}$ by photoassociation with  the  $\sigma^+$
laser instead of the $\pi$ laser\cite{photoasso}, and
then the three atoms form the
  $^2P_{3/2} +^2S_{1/2}+^2S_{1/2}$ trimer.
By modulating the relative depth of the $\sigma^{\pm}$ potential,
 the effective potential is obtained as illustrated by a dashed line in the figures.
The trimer tunnels to site $i_m$,
 and then emission spontaneously occurs
   towards the $^2S_{1/2}+^2S_{1/2} +^2S_{1/2}$ atoms on site $i_m$.
Similarly, just by ``feeling'' the local maximum of the potential of the $\sigma^-$ laser,
as illustrated by a solid line, the trimer tunnels into site $i_a+1$.
Fi\-gure \ref{ex}(b) shows that the atoms in the trimer can possibly separate and
 tunnel into sites with different directions.
However, the  probability of a separation is smal\-ler than the trimer tunneling.

\subsection{Theoretical derivation}
The interactions of the whole system  are of three types: atom-atom, atom-trimer, and trimer-trimer interactions.
Since the trimers are virtually excited states and disappear quickly,
we neglect the third type of interaction.
The interactions can be expressed as field ope\-rators,
and those ope\-rators are expanded in terms of creation and  annihilation ope\-rators.
Finally, by  the Schrieffe-Wolf (unitary) transformation   $H_{eff}=e^{-S}He^{S}$ \cite{xfzhou,je},
the effective Hamiltonian of the system
is found to be:
\be
H_{eff}=H_1+H_{2}+H_{3},
\label{ham}
\ee
with
\be H_1= -\sum_{\la ij \ra}\left[\left(t+t_1 n_i^2 +t_1 n_j^2 \right)a_i^{\dagger}a_j +H.c.\right],
\ee
where, $t$ is the atom tunneling energy\cite{hopt1,hopt2},
$t_1$ is the density-dependent atom tunneling energy ( similar terms have been discussed in pre\-vious work\cite{ddt1,ddt2}),
$H_1$ completely represents the atom tunneling term and $\la ij \ra$ denotes the nearest neighborhood sites.
The second term $H_2$  exhibits the pair and trimer tunneling of atoms  and is given by:
\be
H_2=-\sum_{ \la ij \ra}\left[\left(t_2n_i+t_2n_j\right)a_i^{\dagger2}a_j^2+J a_i^{\dagger3}a_j^3+H.c.\right],
\ee
where $t_2$ describes the density-dependent pair tunneling element, and $J$ means the trimer tunneling element.
The on-site interaction is:
\be H_{3}= \sum_i\left[\frac{U}{2}n_i(n_i-1)-\mu n_i+\frac{W}{6}n_i(n_i-1)(n_i-2)\right],
 \ee
where $U$ is the on-site interaction, $W$ is the three-body attractive interaction, and
 $\mu$ is the
chemical potential.
In our model, the parameters can be tuned according to the atom-molecule coupling strength with
\begin{equation*}
\begin{split}
 J&=2\frac{\Omega_1\Omega_2}{\Delta},\\
 t_1&=2\frac{\Omega_1\Omega_4}{\Delta}=2\frac{\Omega_2\Omega_3}{\Delta},\\
 t_2&=2\frac{\Omega_2\Omega_4}{\Delta}=2\frac{\Omega_1\Omega_3}{\Delta},\\
 U&=U_0-\frac{12}{\Delta}(\Omega_1^2+\Omega_2^2 ),\\
 W&=W_0+\frac{12}{\Delta}(\Omega_1^2+\Omega_2^2 ).
\end{split}
\end{equation*}
 In the derivation, we neglected the terms proportional to $\Omega_3 \Omega_4$, which is an
order of magnitude smal\-ler  than $J$.
The coefficients are defined as follows:
\begin{equation*}
\begin{split}
\Omega_{1(2)}&=\Omega\int dx w_m^{*}( x-x_{i_m})w_a( x-x_{i_a(i_a+1)})^3\\
\Omega_{3}&=\Omega\int d x  w_m^{*}( x-x_{i_m})w_a( x-x_{i_a})^{2}w_a( x-x_{i_a+1})\\
\Omega_{4}&=\Omega\int d x w_m^{*}( x-x_{i_m}) w_a( x-x_{i_a})w_a( x-x_{i_a+1})^2,
\end{split}
\end{equation*}
where
the $w_{a(m)}(x)$  are the ground state  Wannier functions for an atom (a trimer) in an optical lattice potential
localized around the position $x$.

\subsection{The  range of the parameters}
\begin{figure}[b]
\includegraphics[width=\columnwidth]{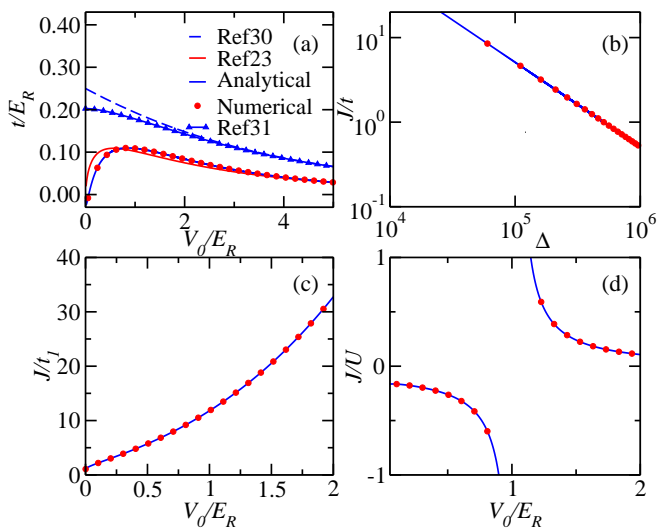}
\caption{(Color online) (a) The {\it exact} integral of the
 hopping matrix $t/E_R$ (solid blue line)  and the previous result\cite{hopt1,hopt2,jaksch}
 shown as function of $V_0/E_R$. Numerical integration results (red solid symbols)
 are also shown to check the analytical integrations.
 Solid triangles are data from Frank\cite{private}. (b) $J/t$ vs $\Delta$, $\Omega_1=10^5 Hz$, $\Omega_2=8.4\times 10^3 Hz$,
 $V_0/E_R=1$.  (c)  $J/t_1$ vs $E_R/V_0$, at  $\Delta=10^5 Hz$, $\Omega_1=10^5 Hz$, $\Omega_2=8.4\times 10^3 Hz$.
 (d) $J/U$ vs $V_0/E_R$, $a_{1D}=1.27\times10^{-12} m$,  $\Omega_1=4\times10^5 Hz$, $\Omega_2=3.36\times 10^4 Hz$.}
\label{gongshi}
\end{figure}

  The ratios between the parameters $J/t$, $J/U$, and $J/t_1$  can be tuned in a wide  range,
$0.5\textless J/t\textless 20$, $-1 \textless J/U \textless 1$ and $0 \textless J/t_1 \textless 35$,
 and will cover the phase transition points\cite{qpt} studied in the following
sections. This will
help us to observe the outcome in the optical lattice.

Figure \ref{gongshi} (a) shows the   hopping-matrix elements
 by three different methods.
Previous work \cite{hopt1,hopt2}   showed the exact solution in the limit
$V_0/E_R\gg1$ from the one dimensional Mathieu equation. The result is
\be
\frac{t}{E_R}=\frac{4}{\sqrt{\pi}} s^{\frac{3}{4}}e^{-2\sqrt{s}},
\label{hop}
\ee
where  the ratio is   $ s=\sqrt{\frac{V_0}{E_R}}$ and the coil energy is $E_R= \frac{\hbar^2k^2}{2m_a}$.
In the present work, through exact symbolic integration from the Maple computer algebra system\cite{maple},
 we get the analytical expression for the hopping element $t/E_R$, and the analytical expression is
\be
\frac{t}{E_R}=\frac{1}{4}  (\pi^2s^2-2s^2-2s) e^{-\frac{\pi^2}{4}s} -\frac{1}{2}s^2e^{-\frac{1}{4} \pi^2s- \frac{1}{s}}
\ee
Fi\-gure \ref{gongshi}(a) shows  that numerical integration is consistent with the  analytical results and
thus both approaches are vindicated.
The result  of Eq.~(\ref{hop}) becomes consistent with the analytical integration
in the regime about $V_0/E_R\ge4$.

The width of the lowest band energy\cite{jaksch} is used to eva\-luate the hopping element $t/E_R$,
from which we know the degree of approximation of the Gaussian Wannier function.
An alternative definition of the hopping element $t/E_R$, which does not require the definition
of Wannier states is a big advantage\cite{private}. The hopping element is defined by\cite{private,frankthesis}
\be
J_{exact}=\frac{1}{N}\sum_k E_k e^{ika},
\label{exact}
\ee
where $N$ is the numbers of the sites, $a$ is the crystal length of the lattice, and $k$ is the wave length
of the particle.
This equation is also  the Fourier
transform of Eq.~(43) of Bloch's review\cite{hopt2}.

The expression of the trimer tunneling $J$ can also be obtained analytically
 by the $\Omega_1$ and $\Omega_3$ and  $\Delta$.
Here
\begin{equation*}
\Omega_1=\Omega\sqrt{1/2\pi}\alpha e^{ -\frac{3}{128}\alpha ^2 \lambda^2 },
\end{equation*}
 and \begin{equation*}\Omega_3=\Omega\sqrt{1/2\pi}\alpha e^{ -\frac{11}{128}\alpha ^2 \lambda^2 }.
\end{equation*}
The ratio of $J/t$ varies from $0.5$ to $20$ as a function of $\Delta$,
as shown in Fig.~\ref{gongshi}(b).
Similarly, the ratio of $J/t_1$ is also given in Fig.~\ref{gongshi}(c).

To obtain the ratio of $J/U$, we should consider the on-site interaction  $U_0$ as
follows \\\be U_0=g_2\int_{-\infty}^{+\infty}|w_a(x)|^4dx=\frac{g_2\alpha}{\sqrt{2\pi}}, \ee
where the coefficient $g_2$ is the coupling constant of the repulsion between two bosons and
$g_2=\frac{2\hbar^2}{m a_s}$ in one dimensional optical lattice\cite{1dlength}.
The ratio $J/U$ is shown in  Fig. \ref{gongshi}(d).

\section{Methods and observables }
\label{sec:method}
 To confirm the results presented in this article,  we use the mean field method (MF), the
 self-improved Quantum Monte-carlo (QMC), and
the density matrix renormalization group (DMRG) method. We compare the results from each for certain parameters.

\subsection{Mean field method}
The MF  method is a very useful and efficient tool to deal with the quantum many body problem,
though the results need to be checked by numerical methods.
Here, we show how to deal with the trimer tunneling, the density dependent tunneling term and the ef\-fects
of temperature in the frame of MF.

Similarly, with  the  decoupling approximation\cite{dec} defined by:
\be a_{i}^{\dag}a_{j}=\la a_{i}^{\dag}\ra a_{j}+a_{i}^{\dag}\la a_{j}\ra - \la a_{i}^{\dag}\ra \la a_{j}\ra, \ee we get
\be a_{i}^{3\dag}a_{j}^3=\la a_{i}^{\dag3}\ra a_{j}^3+a_{i}^{\dag3}\la a_{j}^{3}\ra -\la a_{i}^{\dag3}\ra \la a_{j}^{3}\ra. \ee
To simplify the equations, we let  $\Psi_a=\la  a\ra$, $\Psi_t=\la a^3 \ra$ and $\rho=\la n \ra$.

 For the density dependent terms, we decouple them  in terms of: \be
n_{i}a_{i}^{\dagger2}a_{j}^2=\la n_{i}a_{i}^{\dagger2}\ra a_{j}^2 +  n_{i}a_{i}^{\dagger2}\la a_{j}^2 \ra- \la n_{i}a_{i}^{\dagger2}\ra \la a_{j}^2 \ra
,\ee and
\be
n_{i}^2a_{i}^{\dagger}a_{j}=\la n_{i}^2a_{i}^{\dagger}\ra a_{j} +  n_{i}^2a_{i}^{\dagger}\la a_{j} \ra- \la n_{i}^2a_{i}^{\dagger}\ra \la a_{j} \ra
,\ee where the variational parameters are denoted as
$\Psi_{na}=\la n_{i}^2a_{i}^{\dagger}\ra$,
$ \Psi_ {np} =\la n_{i}a_{i}^{\dagger2}\ra$ and $\Psi_ p=\la a_{j}^2 \ra$ .
We can  deal with $a_{i}^{\dagger2}n_ja_{j}^2$ by a similar method and we only need to pay attention
to the case $\la na^2  \ra \ne \la na^{2\dagger} \ra$.
The work of Ref\cite{mfde} also uses the different variational order parameter for
different density dependent terms.

The Hamiltonian
on the total lattice can be considered as the sum over local ope\-rators on sites
$i$ and $h_i$ and is as follows:
\be
\begin{split}
h_{i}=& -zt\left [ (a_{i}^{\dag}+a_{i})\Psi_a+\Psi_a^2\right] \\
& -zt_p \left[(a_{i}^{3\dag}+a_{i}^3)\Psi_t+ \Psi_t^2\right]+H_3
\end{split}
\ee
where we neglect the density dependent terms and  $z$ is the  coordination number.
The stable solutions of order parameters  are solved iteratively for the ground states.

To explore the temperature ef\-fects\cite{TMF2}, we solve the partition function and the free energy,
$Z=\sum_{k=1}^{n_t+1} e^{-\beta E_k}$ and $F=-\frac{1}{\beta}\ln{Z}$.
For given $U$, $t$, $\mu$, and $T$, the superfluid order parameter
{$\Psi_t$, $\Psi_a$} can be determined by minimizing the free energy, i.e.,
$\frac{\partial^2 F}{\partial  \Psi_a \partial \Psi_t}\biggr|_{U,t,\mu,T}=0$.
The gradient descent algorithm  is used to find the minimum
free energy.
We determine the phase diagrams according to the values in Tab. \ref{t1}.
The compressibility $\kappa$ is an order parameter
for finite temperature phase transitions\cite{kapa}.

\begin{table}[b]
 \caption{
 Values of the order parameters for typical phases.
 }
 \begin{center}

\begin{tabular}{cccccc}
  \hline
  \hline
                ~ & ~ ~ASF~~   & ~~ PSF~~   & ~~ TSF~~  &~ ~MI~~& ~~ NL~~    \\ \hline
  $\Psi_a$       & $\ne0 $  &  0       &   0        &   0 &   0      \\
  $\Psi_p$       & $\ne0 $  & $\ne0 $  &   0    &   0 &   0      \\
  $\Psi_t$       & $\ne0 $  &  0       & $\ne0$ &   0 &   0      \\
  $\kappa$       & $\ne0 $  & $\ne0$   & $\ne0$ &   0 &   $\ne0$ \\
  $\rho_s^{a}$ & $\ne0 $  &  0       &   0    &   0 &   0      \\
  $\rho_s^{t}$ &   0      &  0       & $\ne0$ &   0 &   0      \\
  $C_a$ & $\ne0 $  &  0       &   0    &   0 &   0      \\
  $C_t$ &   0      &  0       & $\ne0$ &   0 &   0      \\
  \hline
  \hline
\end{tabular}
\label{t1}
\end{center}
\end{table}
\subsection{Improved Quantum Monte-carlo method}
To check the results obtained from the MF method,
we simulate the model (\ref{ham})
using the stochastic series expansion (SSE) QMC method\cite{sse}
with the di\-rec\-ted loop update\cite{direcloop}.
The head of a di\-rected loop
carries a creation (annihilation) operator
$a (a^{\dag})$ in the conventional di\-rected loop algorithm for the BH model
with single particle hopping.
To simulate the pair superfluid phase,
the algorithm is improved by allowing the
head of a di\-rected loop to carry a pair \cite{lode,wz2,psf}  of creation (annihilation) ope\-rators
$a^{2\dag}$($a^{2}$). In the present work, to make the atom trimers active, we let the loop heads carry
 trimer creation (annihilation) ope\-rators  $a^{3\dag}$($a^{3}$).

To distinguish the ASF and the TSF states,
we define two types of  superfluid stiffness
$\rho_s^{\alpha}$ as the order
parameter\cite{sfs}:
\begin{equation}
\rho_{s}^{\alpha}=\frac{L^{2-d}\langle W(\alpha)^2\rangle}{2d\beta(9J+t)}.
\label{rho}
\end{equation}
 If $\alpha=t$,
$W(\alpha)$ is the win\-ding number which can be divided without remainders.
If $\alpha=a$,
$W(\alpha)$ is the win\-ding number which cannot  be divided without remainders.

For a TSF phase,  we define the trimer
 superfluid order parameter $\rho_s^{t}>0$ and $\rho_s^{a}=0$.
For an ASF phase, $\rho_s^{a}>0$ and  $\rho_s^{t}=0$.
The factor  `$9$' multiplying  $J$ is due to the hopping of trimer atoms,
or can be understood as the inverse vorticity $\nu=1/3$.
An integer vorticity  $\nu=1$ is found for the ASF phase, and a half vorticity is found for the PSF phase \cite{mfy2,fintemp}.
Now we get a new vorticity $\nu=1/3$ for the TSF phase.
The parameters $d$ and $L$ are respectively the system dimensionality and size, and $\beta$ is the inverse temperature.


\subsection{ The density matrix renormalization group method}
To reiterate, we also use the DMRG method\cite{dmrg1} to
study the system.
Specifically, the DMRG  method is used to impart
the tunneling correlations $C_a=\sum_rC_a(r)/L$ and $C_t=\sum_rC_t(r)/L$,
where   $C_a(r)=\la a_i^{\dagger}a_{i+r}\ra$ and $C_t(r)=\la a_i^{\dagger3}a_{i+r}^3 \ra$.
In our calculations,
we impose periodic  boundary conditions
and restrict the filling factor to $n_{max}=3$.
An  infinite-size algorithm at the chain length of $L= 200$ is used
to ensure a maximum number of possible correlations.

\section{Ground state phase diagram and Trimer superfluid  }
\label{sec:gpt}
As shown in Fig.~\ref{gongshi}(c), for deep lattices, $(J,t,U) \gg t_{1(2)}$.
 The Hamiltonian is reduced to:
\be H=-\sum_{\la ij \ra}(Ja_i^{\dagger3}a_{j}^{3}+ ta_{i}^{\dagger}a_j+H.c.) +H_3
\label{jteq}
\ee
The controllability of the parameters allows us to study the quantum phase transitions
between the ASF, TSF and MI phases.
\subsection{$U=0$, $t=0$ and $\mu/J\ne 0$}
\begin{figure}[htpb]
\includegraphics[width=\columnwidth]{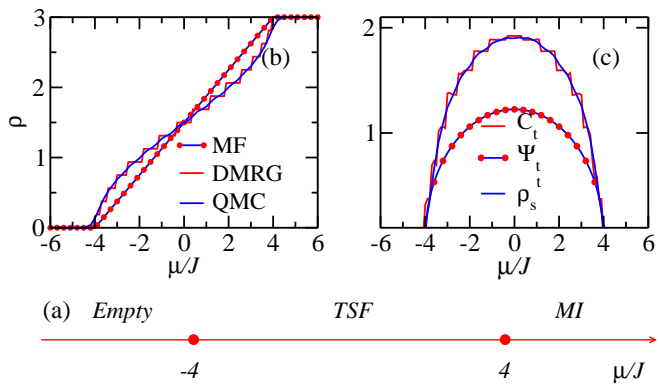}
\caption{(Color online). The parameters are at $U=0$, $t=0$.
(a) Phase diagram  for the model in (\ref{jteq}) which contains empty, TSF, and MI($\rho=3$) phases.
(b) $\rho$ vs $\mu/J$ by the three methods (MF, DMRG and QMC)  for the model in (\ref{jteq}). (c) $\Psi_t$, $C_t$  and $\rho_s^{t}$ vs $\mu/J$.
 }

\label{Jmu}
\end{figure}

Starting with the limit $U=0$ and $t=0$ , the TSF phase will compete with the empty phase and MI($\rho=3$) phase, as shown in Fig.~\ref{Jmu}(a).
To illustrate the competition in more detail, the densities are plotted in the range of  $-6<\mu/J< 6$ by three different methods
in Fig.~\ref{Jmu}(b).
In the  MF frame, we  solve the matrix $h_i$ and then get the minimum energy in the ground state energy
 $E=zJ\Psi^2_t-\frac{3}{2}\mu-\frac{1}{2}\sqrt{9\mu^2+24z^2J^2\Psi^2_t}$.
The location of  the global minimum  $E$ is at $|\Psi_t|= \frac { \sqrt{-6\mu^2+24z^2J^2 }}{4zJ}$.
Substituting  the $\Psi_t$ into $h_i$, we get the exact density  $\rho=\frac{3}{8}(\mu+4)$.
Due to the ignorance of quantum fluctuation, the MF results is a straight line.
 The density from DMRG method looks like  a  staircase due to the finite size effects ($L=16$).
 QMC method gives us the same results with sufficiently lower temperature. If we increase the temperature to $\beta=2$,
the data will change  continuously  due to finite temperature.

Now we understand the transition points between the MI and TSF phases.
For the single particle picture, a trimer emerging on the empty phase will get $-6zJ$
 worth of kinetic energy,  but will be subjected to a
cost of $-3\mu$ worth of potential energy.
For the one dimensional lattice, the coordinates number is
 $z=2$ and consequently  we get  $\mu_c/J=\pm 4$.
If we set  $|\Psi_t|=\frac{3}{4}\sqrt{\frac{16-{\mu_c^2/J}^2}{6}}=0$, the transition point can be made
available again.

Figure  \ref{Jmu}(c)  shows the trimer superfluid stiffness   in more detail.
The transition points between the MI and TSF phases are consistent with those of the MF method.
The maximum  $\rho_s^{t}$  is about half of the maximum  occupation number, which is
similar to the case of the superfluid stiffness discussed before\cite{magloop}.
The data converges well for different sizes, so we choose $L=16$, without any loss of generality.

\subsection{$t=0$, $\mu/U \ne 0$ and $J/U\ne 0$}
\begin{figure}[htpb]
\includegraphics[width=\columnwidth]{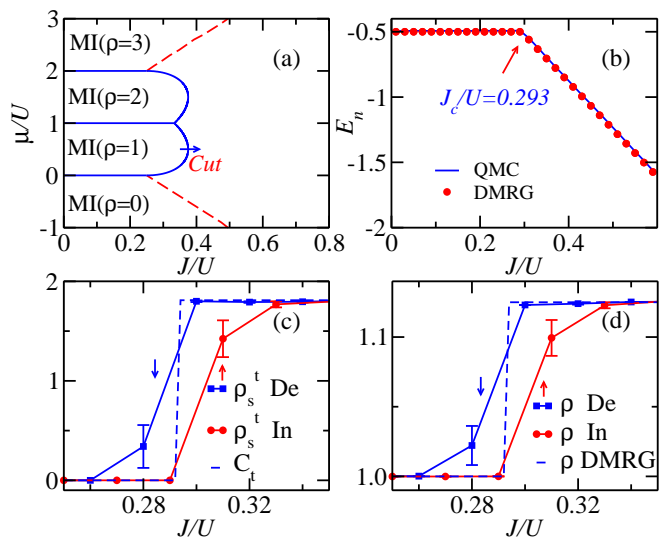}
\caption{(Color online). (a) Phase diagram ($\mu/U$,$J/U$)  of the model (\ref{jteq})  for
$t=0$.
The red dashed line and blue solid line are the continuous and first order transitions,
respectively.
(b) Energies per site $E_n$ along the cut $\mu/U=0.5$ at the size $L=8$.
The red circles (DMRG) and blue lines (QMC)  are consistent and the phase transition point  $J_c/U=0.293$.
 (c) Hysteresis loop of the $\rho_s^t$ vs $J/U$  along the cut $\mu/U=0.5$, and it is a signal of first order transition.
 The QMC data are obtained  in a closed loop  by increasing and decrea\-sing  $J/U$. (see text).
 $C_t$ is also shown   to confirm the transition point.
 (d) Hysteresis loop of $\rho$ vs $J/U$ and $\rho$ behaviors similar to $\rho_s^t$.
 }
\label{JU}
\end{figure}

Fi\-gure \ref{JU} (a) shows that,  in the presence of an on-site repulsive interaction, MI($\rho=1$) and MI($\rho=2$) emerge between the TSF phases.
The exact boundary of the MI-TSF phases can be derived in the MF frame.
By solving the Hamiltonian matrix, four energy eigenvalues are obtained.
The lowest one for $0<\mu<1$ is:
\[E=zJ\Psi^2_t-\frac{3}{2}(\mu-U)-\frac{1}{2}\sqrt{9(\mu-U)^2+24z^2J^2\Psi^2_t} ~.\]
Letting $\frac{\partial E}{\partial \Psi_t}=0$, the superfluid order parameter is
$|\Psi_t|=\frac{\sqrt{6}}{2}\sqrt{1-\frac{(\mu-U)^2}{4z^2J^2}}$,  for $1-\frac{(\mu-U)^2}{4z^2J^2} \geq 0$.
By substituting $|\Psi_t|$ into the energy expression, we get:
\[E_3=\frac{3(\mu-U)^2}{8zJ}-\frac{3}{2}(\mu-U)-\frac{3}{2zJ} ~.
\]
Along the boundaries of the MI-TSF phases, the energy of both phases should be equal ($E_3=-\mu $).
The boundary is obtained as  $J=\frac{3U-\mu}{12}+\frac{\sqrt{3\mu U-2\mu^2}}{6}$.
For example, if  $\mu/U=0$,  then  $J/U=0.25$, which is consistent with the previous  expression for $J/U$.

To confirm and check the MF phase diagram,  the energies are plotted along the cut $\mu/U=0.5$
by both DMRG and QMC methods and the results are consistent again.
We find the energy makes a corner at $J_c/U=0.293$, smal\-ler than the
MF phase transition point at $J_c/U=0.375$, which is in the range shown
 in Fig.~\ref{gongshi}(c).

Figure \ref{JU}(c) and (d)  show  two  hysteresis loops for
superfluid stiffness and density,
 which are the  signals of  first order transitions.
In pre\-vious work\cite{magloop}, one of us  found a multi-hysteric loop phenomena and located
 the phase  transition point successfully, by
increasing and decrea\-sing the parameters in a cycle.
In the present work, we use the same method. Firstly,
 we obtain the  physical quantities  $\rho$ and  $\rho_s^t$, by increasing
 the variable $J/U$. When we decrease $J/U$,  a closed loop takes form, as shown in Fig.~\ref{JU}(c) and (d).
 In the simulation, we store the last configuration, and then input it  as the next initial configuration.
We find that the closed loops of $\rho$ and $\rho_s^t$ behavior in a similar
fashion.
The correlation $C_t$  is also consistent with $\rho_s^t$.

\subsection{$J/t \ne 0$, $\mu/t \ne 0$  and $U = 0$ }
\begin{figure}[htpb]
\includegraphics[width=\columnwidth]{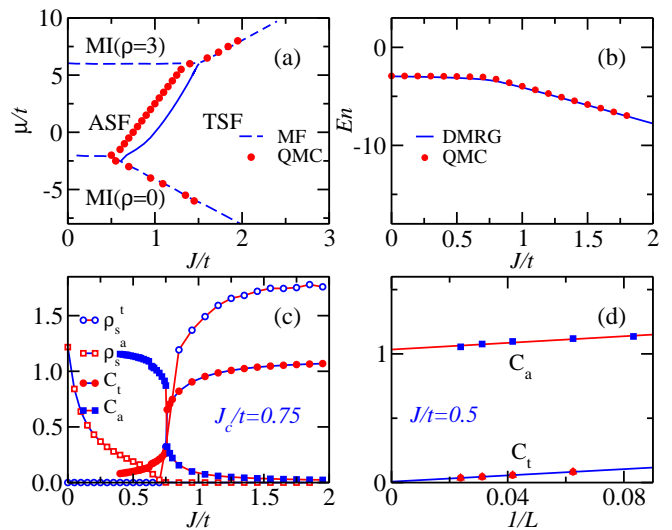}
\caption{(Color online). (a) Phase diagram    $(J/t, \mu/t)$ of the model (\ref{jteq}) for $U=0$.
The blue dashed lines and blue solid line are the continuous and first order transitions,
respectively. The red solid circles are the QMC results.
 (b) Energy per site $E_n$ vs $J/t$ by DMRG and QMC methods  along the cut $\mu/t=0$ at the size $L=24$.
 (c) $C_t$, $C_a$, $\rho_s^{a}$ and $\rho_s^{t}$ vs $J/t$.
 (d) The finite size scalings of  $C_t$, $C_a$ in the ASF phase at $J/t=0.5$. }
\label{Jt}
\end{figure}

With both $J$ and $t$ are nonzero, it  is difficult to get the analytical boundary lines.
Here we provide the numerical results. Figure \ref{Jt}(a) shows the phase diagram in the plane $(J/t, \mu/t)$,
 containing the  ASF, TSF and MI phases.
The system exhibits an ASF phase for small $J/t$, a TSF phase for lar\-ger $J/t$, an empty phase for
 small $\mu/t$ and a MI($\rho=3$) for lar\-ger $\mu/t$.
The ASF-TSF phase transition is clearly first order.
With $J=0$, the systems undergo a MI($\rho=0$) to ASF phase transition at $\mu/t=-2$    and a MI $(\rho=3)$ to the ASF phase transition at $\mu/t=6$, which can  be understood by a single particle picture.

Figure \ref{Jt}(b) compares the energies per site $E_n$ along the cut $\mu/t=0$.
The energies from the different methods are well consistent with each other. They keep a constant in the
ASF phase for $J/t<0.75$ and then go down in a  straight line for $J/t>0.75$.

To illustrate the ASF-TSF phase transition in more detail.
We shows $C_t$, $C_a$, $\rho_s^{a}$ and $\rho_s^{t}$ vs $J/t$.
In the range $0<J/t<0.75$, we find $\rho_s^{a}$  decreases as  $J/t$ increases, and
$\rho_s^{a}$ tends to  zero  continuously  at the phase transition point  $J_c/t = 0.75$.
$\rho_s^{t}$  jumps  to nonzero values at the same location as $\rho_s^{a}$ changes.
These phenomenon clearly proves that the ASF-TSF phase transition is indeed
of first order instead of a weak first order transition\cite{wz1,wz2}.
In the whole range, we also see the sharp jumps of $C_t$ and $C_a$ as $J/t$
increases. In the ASF phase, $C_t=0$  while $C_a \ne 0$ in the thermal dynamical limit, as shown in Fig.~\ref{Jt}(d).
 In the TSF phase, $C_t\ne 0$  while $C_a=0$, which is not shown.

\subsection{Winding numbers and Correlation}
\begin{figure}[t]
\includegraphics[width=0.9\columnwidth]{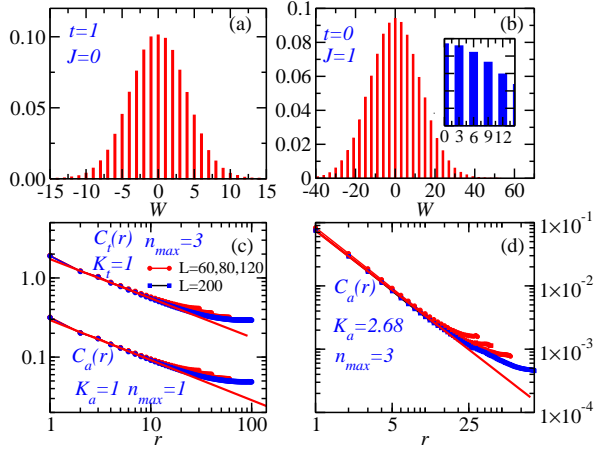}
\caption{ Histograms of the win\-ding numbers $W$ for (a) $t=1$ is the only nonzero parameter and $W$ varies in units of unity. (b)
 $J=1$ is the only nonzero parameter and $W$ occurs  as  $0,\pm3, \pm6, \pm9,...$.
(c) The power law decay of  the $C_t(r)$ in the TSF phase
with $n_{max}=3$, and the $C_a(r)$ in the ASF phase with $n_{max}=1$. $K_{a(t)}=1$.
(d) The power law decay of the correlation $C_a(r)$ for the softcore bosons. $K_{a}=2.68$.}
\label{hist}
\end{figure}

In Figs.~\ref{hist} (a) and (b) , we give a typical win\-ding number distribution  in the ASF and TSF phase.
Clearly, both of them are  represented
by a Gaussian distribution.
For the ASF phase, we set $t=1$ and all  the other parameters are zero.  The win\-ding numbers vary
in  units of one, such as $0$, $\pm 1$, $\pm 2$, $\cdots$.
However, for the TSF phase, we set $J=1$ and all  the other parameters are zero. The win\-ding numbers only occur as numbers
which  can be divided by $3$
without any remainders, such as  $0$, $\pm 3$, $\pm 6$, $\cdots$. This can be
understood  as   three atoms are bound together tightly and   move in space simultaneously.

To find the relationship between the ASF phase and the TSF phase,
in addition to  the distribution
of win\-ding numbers, we also show the correlation between trimers and holes.
We let the size of the system be as large as possible i.e. $L=60$, $80$, $120$, and $200$.

In Fig.~\ref{hist}(c), we plot the trimer correlation $C_t(r)$, and  the  correlation satisfies the power law decay:
\be
C_t(r)= a_t^0r^{-\frac{K_t}{2}}
\ee
where the fitted result is $K_t=1$.
For the ASF phase in hardcore bosons  (the on-site maximum number is $n_{max}=1$),
the correlation satisfies:
\be
C_a(r)= a_a^0r^{-\frac{K_a}{2}},
\ee
where the fitted exponent  is also $K_a=1$.
The two types of correlation decay with a same exponent. This is
because the TSF state can be mapped to the  ASF state, if we consider the trimer as a whole  molecule.
At the same time, the above two cases are   non-interactive systems\cite{colex}.
However, if $n_{max}=3$,
then  $K_a=2.68$ which is lar\-ger than $K_a=1$, as shown
in Fig.~\ref{hist}(d). In this case, the atom  tunneling will be  hindered by the additional bosons.
Then the correlation will decay more quickly than the cases for $K_{a(t)}=1$.

\section{Effects of density dependent tunneling}
\label{sec:dependent}
Besides the atom hopping and trimer hopping terms,
the occupation dependent terms with
$H_{n1}=-t_1\left(n_i^2+n_j^2\right)a_i^{\dagger}a_j$
and $H_{n2}=-t_2\left(n_i+n_j\right)a_i^{\dagger2}a_j^2 $ ,
are easily overlooked, as they are smal\-ler than $J$ and $t$ in a wide range of parameters.
Ref\cite{xfzhou, ddt1,ddt2} only mentioned the similar terms without
the detailed  discussion of the effects of $H_{n1}$ and $H_{n2}$ to the ASF-TSF phase transition.
However, $t_1$ and $t_2$ are coupled with density ope\-rators, which will  change the phase diagram  sufficiently.
So it is necessary to explore the competition between $t_1(t_2)$, $J$ and $\mu$.
However, experimentally, as discussed in Sec.~\ref{sec:model}, the parameter $t_1$ equals $t_2$
and cannot be tuned independently and        consequently we define $t_{12}=(t_1+t_2)/2$ to study
 the  physical consequences of having both $t_1$ and $t_2$ being nonzero.

\begin{figure}[t]
 \includegraphics[width=0.9\columnwidth]{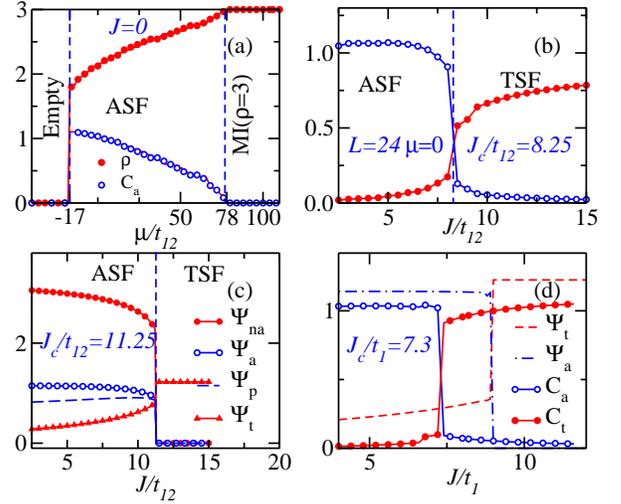}
 \caption{ (Color online) (a) $\rho$ (solid circles)  and $C_a$ (empty circles)  vs $\mu/t_{12}$ with $J=0$.
 The phase transition point is  $\mu_c/t_{12} =-17$ between the empty phase and the ASF phase.
  (b)   $C_t$ and $C_a$  vs $J/t_{12}$ with $\mu=0$.
  The  ASF-TSF  phase transition point is  at  $8.25$ by the DMRG method.
 (c) MF results of quantities  $\Psi_p$, $\Psi_a$, $\Psi_{na}$ and $\Psi_t$ as function of   $J/t_{12}$.
 The  MF phase transition point is at $J/t_{12}=11.25$.
(d) MF and DMRG results of quantities  $C_t$, $C_a$, $ \Psi_a$ and $\Psi_t$ as functions of   $J/t_1$.}
\label{dependent}
 \end{figure}

Figure \ref{dependent} (a) shows  $\rho$  vs $\mu/t_{12}$ at  $J=0$. The jump in the densities
 and  the correlation $C_a$ confirm that the  empty -ASF phase transition is first order.
However, the  empty phase to ASF phase transition is continuous for the BH model with $t_{12}=0$.
The accurate phase transition point is available
and is at $\mu_c/t_{12}=-17$.
The single particle picture fails to analyze the  empty-ASF phase transition  points.
Due to the presence of the  term like  $a_i^+n_j^2a_j$,  the system needs at least
two atoms emerging on the site $j$ from the empty to the ASF phase.

Furthermore, by turning on $J$,  we show the density dependent tunneling $t_{12}$  effect to
the  quantum phase transition at $\mu=0$ in Fig.~\ref{dependent} (b).
Compared with $J_c/t=1$ in Fig. \ref{Jt} (a), the phase transition point is $J/t_{12}=8.25$, sufficiently modified by
      the coupled density ope\-rators.
In the presence of the finite size effect, $C_a\ne 0$ in the TSF phase.
 However, we find   $C_a=0$ and $C_t \ne 0$  in the thermal dynamical limit, which is not shown.

Figure \ref{dependent} (c)  shows  the MF results of quantities
  $\Psi_p$, $\Psi_a$, $\Psi_{na}$, and $\Psi_t$ as function of   $J/t_{12}$.
  In the ASF phase, all of the quantities are nonzero while only $\Psi_t$ is nonzero   in the TSF phase.
                       To check which term causes  the first-order ASF-TSF phase transition by $J/t_{12}$,
                        we set $t_2=0$ and only show effects of  $J/t_1$ in Fig.~\ref{dependent} (d).
                        The phase transition point is at $J/t_1=7.3$ and also sufficiently modified by
the coupled density ope\-rators. The MF results are also given, in similar fashion
to the case of $J/t_{12}$.

\section{Finite temperature effects}
\label{sec:finitet}
\begin{figure}[t]
\includegraphics[width=\columnwidth]{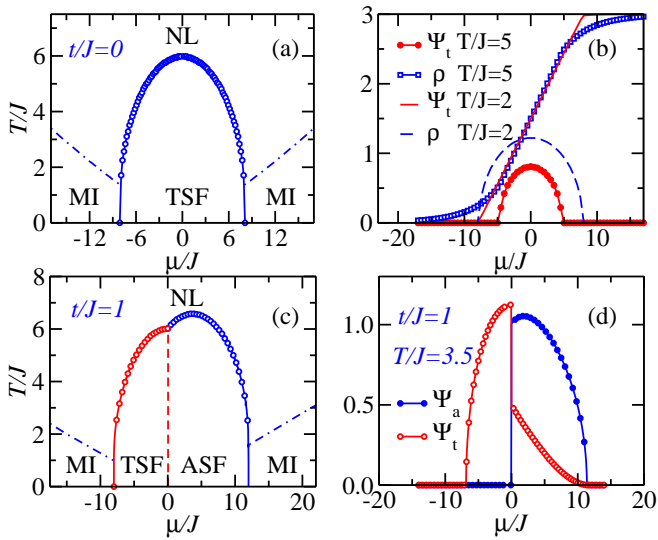}
\caption{ (Color online)
$U = 0$ and $t=0$
(a) Finite temperature MF  phase diagram
($\mu/J$, $T/J$), which contains MI, TSF, NL phase  for  $U = 0$ and $t=0$.
(b) $\Psi_t$ and $\rho$ vs $\mu/J$  for  $T/J=2$ and $T/J=5$.
(c) The finite temperature MF phase diagram ($\mu/J$,$T/J$).
(d) $\Psi_t$ and $\Psi_a$ for $T/J=3.5$.}
\label{wendu}
\end{figure}

Since there is no thermal  Berezinskii-Kosterlitz-Thouless(BKT) phase transition for the 1-D classical or quantum system,
we choose a two dimensional system\cite{kt} as a candidate to
study the thermal melting of the TSF.

The phase diagram in both the MF  frame are drawn.
In Fig. \ref{wendu} (a), on both sides, the MI-normal liquid(NL) boundary lines are linear.
As shown in pre\-vious work\cite{kapa},  $\kappa = 0$ in the MI phase.
Since there is no long range order for both the MI and NL phases,
the va\-lue $\kappa$ updates without any singularity into a nonzero va\-lue in the NL phase at finite tempe\-rature.
Therefore, we use the criterion
 $\kappa = 0.002$ to obtain the MI-NL crossover boundary lines.

The TSF phases emerges in a leaf shaped region, for lower tempe\-ratures, in the range of $-8<\mu/J<8$.
The boundaries between MI-TSF phases are consistent with the analysis in the zero temperature limit.
The transition temperatures of the TSF-NL phases varies with different chemical potentials.
At $\mu=0$, the TSF-NL transition point is $T/J=6$.
In Fig.~\ref{wendu}(b), we plot density $\rho$ and correlation $\psi_t$ as function of $\mu/J$
with different temperatures $T/J=2$ and $T/J=5$.

Figure \ref{wendu}(c) shows the finite temperature phase diagram,
where both $t$ and $J$ are nonzero.
We find MI,TSF, ASF phases at lower
temperatures, and NL phase for higher temperatures. There is no obvious difference
between the boundary lines of TSF-NL and the ASF-NL phases.
Therefore, the cold-atom experiment can observe the TSF phase in  the temperatures just  as the ASF phase.
In Fig.~\ref{wendu}(d), we describe $\Psi_t$ and $\Psi_a$ in more detail.
At $T/J=3.5$, we scan $\mu/J$, and find the TSF phase in the range
$-6.8<\mu/J<0$ and the ASF phase in the range $0<\mu/J<11.6$ .
The jump of the order parameters  $\Psi_t$ and $\Psi_a$  clearly demonstrates
that the phase transition between ASF-TSF phases is of first order, even at finite temperatures.

\section{Discussion and conclusion}
\label{sec:discon}

The ASF phase is of type $ \la a \ra \ne 0$,  $\la a^3 \ra \ne 0$  and  $Z_2$ broken symmetry,
while the TSF phase is of type  $\la a^3 \ra \ne 0$  and $U(1)$ broken symmetry\cite{sym}.
According to the Landau-Ginzburg-Wilson paradigm,
different broken symmetry patterns make the ASF-TSF quantum phase transition first order.
On the other hand, the $Z_2$ broken symmetry corresponds to an Ising phase transition.
However, by  the Coleman-Weinberg mechanism\cite{mecha} and our calculation,
 the ASF-TSF phase transition becomes a first-order one driven by the  large quantum  fluctuation.

At finite temperatures, we choose a two-dimensional lattice as a platform to study the
temperature effects. We provided the MF  results of the transition points of
the thermal melting of the TSF phase.  By the MF method, the ASF-TSF phase transition
is first order at $T>0$. Further careful checking is needed by other methods.

In conclusion, we derived a new  BH model with trimer tunneling and two types of
density dependent tunneling.
We improved the QMC algorithm for simulation of trimer tunneling.
We studied  the ground state
quantum phase diagram in a one-dimensional lattice by three different methods, which agree with each other in applicable regimes.
The parameters to be realized were also discussed, and various interesting phase transitions were studied.
We found the first-order  empty-ASF phase phase transition in the presence of the density dependent tunneling terms.
Our results will be helpful in guiding the  optical experimentalists  to realize the trimer superfluid.
\acknowledgments
We thank  F. Z. Liu for his help during preparing the manuscript and
the useful suggestions from  N. Prokof'ev, X. F. Zhou and B. Huang.
We thank  Frank Deuretzbacher for the data and understanding of the hopping element $J_{exact}$.
This work is supported  by the NSFC under Grant No.11305113, No.11247251,  No.11204204,
No.11204201 and  the project GDW201400042  for   high-end foreign experts.

\appendix
\section{The  hopping $t/E_R$ for the deep well}
Figure \ref{gongshi} (a) shows  the atom tunneling amplitude $t$, both numerically and analytically, which is defined
by
 \be t=-\int_{-\infty}^{+\infty}w_a(x-x_i)H(x)w_a^*(x-x_j)dx, \label{eq:wannier} \ee where $H(x)=(\frac{-\hbar^2}{2m}\frac{d^2}{dx^2}+V(x))$
is the Hamiltonian of the system. Here the  optical potential is $V(x)=V_0 sin^2(kx)$, where
the wave number $k=2\pi/\lambda$,  $\lambda=500nm$.
The Wannier functions of an atom (or molecule) are of  Gaussian type:
 \be w_{a(m)}(x)=\left[\alpha^2/\pi\right]^{\frac{1}{4}} e^{-\frac{\alpha^2x^2}{2}},
\ee
and the characteristic  length\cite{opticallattice} is :
\be
 \alpha^{-1}= \left(\hbar^2 2m_{a(m)}V_0/k^2 \right)^{1/4}. \ee
Previous work \cite{hopt1,hopt2}   showed the exact solution in the limit
$V_0/E_R\gg1$ from the one dimensional Mathieu equation.
In the present work, through exact symbolic integration from the Maple computer algebra system\cite{maple}, we get the analytical expression for the hopping matrix element:
\be
\frac{t}{E_R}=\frac{1}{4}  (\pi^2s^2-2s^2-2s) e^{-\frac{\pi^2}{4}s} -\frac{1}{2}s^2e^{-\frac{1}{4} \pi^2s- \frac{1}{s}}
\ee
for the ratio   $ s=\sqrt{\frac{V_0}{E_R}}$ and the coil energy $E_R= \frac{\hbar^2k^2}{2m_a}$.
Note that in this integral, we set $x_i=-\lambda/4$ and $x_j=\lambda/4$. The solution is also based on
the approximated Wannier function.
\section{The  hopping $t/E_R$ for the wells with any depth}

According to Frank Deuretzbacher\cite{private},  Eq.~(\ref{exact})  can be simplified to:
\be
J_{exact}=- \int_0^1 E_{\kappa} \cos(\pi \kappa) d \kappa,
\label{exact2}
\ee
where $\kappa = \frac{ka}{\pi}$.
$E_{\kappa}$ can be obtained by solving the one-dimensional Mathieu function, and  the expression is
\be
\frac{E_k}{E_R}=a(\kappa,-\frac{V_0}{4E_R})-\frac{V_0}{2E_R}.
\label{ek}
\ee
In the equation above, the function ``a'' is the \verb+MathieuCharacteristicA+  function in Mathematica or
\verb+MathieuA+ in Maple.
Substituting   Eq.~(\ref{ek}) into Eq.~(\ref{exact2}), the  numerical result is available as shown in Fig.~\ref{gongshi}(a).

This result is simply the outcome of having the Hamiltonian operate on the eigenfunction, getting the eigenvalue ``a''
and the simply integrating over the product of Mathieu functions which
yields unity because they are normalized.  To reiterate,
the Mathieu ``a'' function is, apart from an offset, the
energy of a Bloch state (Eq. (\ref{ek})). So, inserting it into Eq. (\ref{eq:wannier}),
the usual definition of $J$,  the formula that connects the
Wannier functions to the Bloch functions (i.e. the discrete Fourier
transform of the Bloch functions),  then using the fact that the Mathieu ``a''
functions are the eigenenergies of the Bloch functions, and from the orthogonality
and the normalization of the Bloch functions, Eq.~(\ref{exact}) is finally obtained.

\end{document}